\documentclass[pre,aps,twocolumn,nofootinbib]{revtex4-1}
\usepackage[latin1]{inputenc}
\usepackage[english]{babel}
\usepackage{graphicx}
\usepackage{color}
\usepackage{amsmath}
\usepackage{amssymb}
\usepackage{epstopdf}
\usepackage{multirow}
\usepackage{caption}

\begin{document}

\title{\bf SAdS black holes and spacetime atoms: a heuristic approach}
\author{A. F. Vargas}
\email{af.vargas1540@uniandes.edu.co}
\author{E. Contreras}
\thanks{On leave from Universidad Central de Venezuela.}
\email{ej.contreras@uniandes.edu.co} 
\author{P. Bargue\~no}
\email{p.bargueno@uniandes.edu.co}
\affiliation{Departamento de F\'{\i}sica,
Universidad de los Andes, Apartado A\'ereo {\it 4976}, Bogot\'a, Distrito Capital, Colombia}
\begin{abstract}
In this work, both extended phase space and holographic equipartition approaches are employed to develop an exact Van der Waals 
description of non--rotating $D=4$ SAdS black holes as an ensemble of spacetime atoms. 
After a possible microscopical interaction model is introduced, 
statistical mechanics techniques, with certain heuristic gravitational constraints, are used to derive the equation of state 
and the Bekenstein--Hawking entropy.
The procedure is generalized to the charged $D=4$ SAdS black hole and to arbitrary $D\ge 3$ dimensions for the uncharged cases. 
\end{abstract}

\maketitle


\section{Introduction} 

Gravity could be an emergent phenomenon as firstly proposed by
Sakharov \cite{Sakharov1968}, concreted by Jacobson \cite{Jacobson1995} and generalized significantly
by Padmanabhan \cite{Padma2010a}. The realization of this idea is expected to be found within black hole (BH) thermodynamics,
which has been a cornerstone in the search for a full theory of quantum gravity. Along the last years, 
the extended phase space (EPS) \cite{Kastor2009,Dolan2011,Dolan2011bis,Cvetic2011,Dolan2012,Kubiznak2015,Mann2016,Dolan2015} and {\it spacetime atoms} (STA) \cite{Padma2010bis,Padma2010,Padma2010tres,Padma2015} approaches have been implemented to further understand the
nature of gravitational phenomena. The first one, coined BH chemistry \cite{Kubiznak2012}, concerned with macroscopical properties of BHs as 
thermodynamical systems and has been succesfully applied to numerous BHs with emphasis on their particular equations
of state (eos) and phase transitions \cite{Altamirano2014,Kubiznak2016} while the second one deals with a possible microscopical description of gravity due to
 hypothetical spacetime atoms that count the degrees of freedom (dof) through a holographic equipartition law \cite{Padma2010}.

Specifically, the EPS approach has drawn sigficant attention by means of Van der Waals (VdW)--like eos 
for different AdS BHs \cite{Kubiznak2012,Altamirano2014,Gunasekaran2012}. In view of these developments, one step further, using the VdW--like eos, 
has been recently taken in Ref. \cite{Wei2015}, where the possible microscopic structure of a charged AdS BH has been investigated from 
the thermodynamic viewpoint of a phase transition. Even more, in Ref. \cite{Wei2015}, a number density of the BH {\it molecules}  
has been introduced to measure the microscopic dof of the BH. 

Therefore, both EPS and STA approaches include some microscopical ingredient of gravity, namely BH molecules or spacetime atoms.
Consequently, the quest for a statistical treatment, based on these dof, appears as a interesting open problem which could
shed light on the emergence of AdS BH thermodynamics.

We will like to point out that previous works \cite{Ghosh2014,Asin2015} have also considered models of certain dof in order
to obtain BH thermodynamic quantities from a statistical approach within the context of loop quantum
gravity (LQG). Specifically, the authors of Refs. \cite{Ghosh2014,Asin2015} considered an indistinguishable gas of punctures
 obeying Maxwell--Boltzmann statistics. In this sense, these punctures would play the role of either spacetime atoms or BH molecules.

Following these ideas, in this work we show that the VdW--like eos for AdS BHs, which appear in the EPS approach, 
can be derived from the holographic equipartition for STAs. 
In addition, both the VdW--like eos (which turns to be equivalent to 
the Smarr law), together with the Bekenstein--Hawking entropy for AdS BHs in $D\ge 3$, are recovered from a 
particular statistical treatment developed for these STAs. In this sense, BH thermodynamics can be shown to emerge from 
the dof described by STAs.

This work is organized as follows. In Sect. \ref{from}, we show that the VdW--like eos is derived from the STAs approach.
In addition, the thermodynamical system of interest is properly defined. Then, in Sect. \ref{secIII}, 
we propose a {\it dual} system of STAs and we give a microscopical interpretation for the previous eos. Sect. \ref{secIV} 
develops the statistical treatment of these STAs to obtain the known $D=4$ 
AdS BH thermodynamics, which is generalized to $D\ge 3$ dimension in Sect. \ref{secV}. Finally, some crucial points regarding
the interpretation of the statistical treatment are discussed in Sect. \ref{secVI}, followed by the concluding remarks
presented in Sect. \ref{secVII}.

\section{From SAdS to Smarr/VdW}
\label{from}
One more step towards an emergent perspective for gravity has been established by the so called
holographic energy equipartition \cite{Padma2010,Padma2010bis} which, for a static spacetime, reads ($\hbar = G = c  = 1$)
\begin{equation}
\label{Padmaequi}
E\equiv \int_{\mathcal{V}} d^{3}x\sqrt{h}\rho_{K}=\frac{1}{2}\int_{\partial \mathcal{V}} d^{2}x\sqrt{\sigma} \; T,
\end{equation}
where $\mathcal{V}$ is a three--volume bounded by $\partial \mathcal{V}$, with induced metrics whose determinants are given by $h$ and
$\sigma$, respectively. In Eq. (\ref{Padmaequi}), $T$ stands for the local Hawking temperature measured by an observer
at rest in this spacetime and $\rho_{K}$ is the corresponding Komar energy--density. Following \cite{Padma2010,Padma2010bis}, we can attribute $\Delta N=d^{2}x\sqrt{\sigma}$
microscopic dof to an element of area (in the units here used). 
 In the particular case of dealing with an
event horizon corresponding to $\partial \mathcal{V}$, Eq. (\ref{Padmaequi}) can be written as $E = \frac{1}{2}N\, T$.

In the specific case of Petrov--type D solutions, working with the Kinnersley tetrad \cite{Kinnethesis}, 
the volume term in Eq. (\ref{Padmaequi})
can be written as $\left(\frac{R}{12}-\Psi_{2} \right)r^3$. Therefore, Eq. (\ref{Padmaequi}) reads
\begin{equation}
\label{master}
E = \frac{1}{2}N\, T= \left(\frac{R}{12}-\Psi_{2} \right)r_{+}^3,
\end{equation}
where the horizon is located at $r_{+}$, $R$ is the scalar curvature and
$\Psi_{2}$ is the only non--vanishing Weyl scalar for Petrov--type D solutions \cite{Chandrabook}.
On one hand, we note that, as pointed out by Padmanabhan \cite{Padma2010,Padma2010bis}, the first part
of Eq. (\ref{master}) describes the energy equipartition at temperature $T$ of $N = A$ particles with one
degree of freedom. The second part, on the other hand, can be used to construct an eos for $N$ particles. 

For a $3+1$-SAdS BH, $\Psi_{2}=-\frac{M}{r_{+}^{3}}$, where $M$ is the mass of the BH and $R = 4\lambda$. Therefore, after eliminating $M$ 
by using $1-\frac{2M}{r_{+}}-\frac{\lambda}{3}r_{+}^2\nobreak =0$, Eq. (\ref{master}) can be written as 
%
\begin{equation}
\label{state1}
T = \frac{\tilde v^{-1/3}}{4\pi}+ 2 P \tilde  v^{1/3},
\end{equation}
where,  following \cite{Kastor2009, Dolan2011,Cvetic2011,Dolan2011bis},  we have introduced the notation 
$\tilde v =\frac{3}{4\pi}V_{t}$, $V_{t}=\frac{4}{3}\pi r_{+}^3$  and $P=-\frac{\lambda}{8 \pi}$.
We note that Eq. (\ref{state1}) has been obtained in the past by several authors
\cite{Dolan2011,Kubiznak2016} without invoking Eq. (\ref{master}) 
but rather by obtaining the temperature for the BH from the surface gravity and identifying term by term, with the definitions here provided. In fact, this equation
has been studied extensively within the {\it extended phase space} approach to BH thermodynamics (for a comprehensive review on these developments see \cite{Kubiznak2016} and references therein).

Even more, by doing the substitution 
$\tilde v^{1/3} = V_{t}/2\bar N$, as defined in \cite{Kubiznak2012},
Eq. (\ref{state1}) can be recast as
\begin{equation}
\label{vdW}
P = \frac{\bar{N}T}{V_{t}}-\frac{1}{2 \pi}\frac{\bar{N}^2}{V^2_{t}}, 
\end{equation}
where $\bar{N} = N/6$. This factor of $6$ has been introduced in order for the pressure and $V_{t}$ to be conjugates. Note that the same numerical
factor has been recently considered in Ref. \cite{Wei2015} with the aim of giving to the physically relevant parameter, {\it i.e.}, the specific volume, 
$v \vert_{r+} =V_{t}/\bar N \vert_{r+} =2\,l_{p}^2\, r_{+}$\footnote{Here we have explicitly restored the Planck length for clarity} \cite{Kubiznak2012}, an interpretation as a particle density. We note that Eq. (\ref{vdW}) is unique in the sense it follows 
directly from the definition of the specific volume, $v$, within the EPS approach \cite{Kubiznak2016}.  

We note the exact (but {\it formal})
correspondence with a VdW gas of $\bar N$ particles inside a volume $V_{t}$ in equilibrium at the Hawking temperature. We will like to point out that all the thermodynamical variables only depend
on the horizon radius, $r_{+}$. However, we are going to consider them as independent variables in Eq. (\ref{vdW}). 
This is what we mean by {\it formal}. Now well, a usual VdW gas is described in terms of two parameters: $a$ and $b$, which correspond to interactions and covolume terms, respectively. In this
case, the VdW eos is written as $P=\frac{\bar NT}{V_{t}-\bar N b}-\frac{a \bar N^2}{V_{t}^2}$ 
(for an account of the VdW gas see, for example, \cite{Pathriabook}). However, in our case, although
the corresponding parameters are $a= \frac{1}{2\pi}$ and $b=0$, their interpretation differ from the standard VdW gas, as we
will show in the next section.

Even more, when used carefully, 
Eq. (\ref{vdW}) can be employed to give at least an heuristic insight into the microscopical
nature of gravity. In any case, we  emphasize that both Eqs. (\ref{state1}) and (\ref{vdW}) are {\it exact} consequences
of semiclassical general relativity. 
Even more, once the entropy has been taken to be $S=\frac{N}{4}$, it can be seen that Eqs. (\ref{master}) and (\ref{vdW}) 
are nothing but different expressions for the Smarr law for SAdS BHs, which reads $2 TS=2P V_{t}+M$. Given these observations, we shall propose a particular
microscopical model which reproduces both the Eq. (\ref{vdW}) and the Bekenstein--Hawking entropy.

\section{An interaction model}
\label{secIII}

We begin by proposing a {\it dual} system in which we have $\bar{N}$ particles living inside a 
spherical container of fixed volume, $V_{t}$, whose frontier can be thought as 
located at a radius $r_{+}$. Furthermore, these particles have a 
VdW-gas eos given by Eq. (\ref{vdW}). As we know, all the gravitational variables $(\bar{N}, V_{t}, T, ...)$ are functions of the 
radius $r_{+}$, and as such they can have a non-trivial dependence on each other. To avoid this problem, within our {\it dual} 
system, we will regard them as independent variables when performing the calculations and only at the end of them they will 
be ``mapped" to their gravitational counterpart, which depends on $r_{+}$, following the definitions given in Sect. \ref{from}. 
To regard the statistical variables $(\bar{N}, V_{t}, T, ...)$ as independent will be named here as {\it the independence condition}.

In the usual theory, the parameter $a$ is related to the attractive part of the potential which binds the particles of the fluid. 
However, here we can not translate this into the gravitational theory because our starting point, Eq. (\ref{master}), does not 
account for interactions between the $\bar{N}$ particles. 
Therefore, the natural question is: where does this parameter comes from?

We must note that the specific value of $a$ is a consequence of the horizon constraint used to remove the mass. 
To see this, consider the SAdS $g_{tt}$ component, which can be written as $g_{tt} = k-\frac{\lambda}{3}r^{2}-\frac{2M}{r}$.
The constant $k$ is commonly known as the sign of the curvature. Previously we took the specific case $k = 1$, which states 
that our event horizon has positive curvature. But in fact there can be cases where $k = 0$ (for flat horizons) and even the 
exotic case of $k = -1$ (for topological BHs) \cite{Mann1997}. With this in mind, the $a$--parameter can be written as 

\begin{equation}
a = \frac{k}{2\pi}.
\label{ak}
\end{equation} 

Now, given the dependence of $a$ to the curvature, 
a good starting point for the interaction model would be a mechanism involving $r_{+}$. To this end, we propose that 
our particles must spend an energy when they arrive to the walls of the container located at $r_{+}$. This can be interpreted in 
the gravitational side as the existence of an event horizon configuration energy. Stated otherwise, in order to have particles 
at $r_{+}$, certain amount of energy must be spent. As such, the following potential energy is proposed 

\begin{equation}
\mathcal{U}(r_{i}) = -\epsilon \delta(r_{i}-r_{+}),
\label{ansatz}
\end{equation} 

where $r_{i}$ is the position of the $i$--th particle and $\epsilon$ is a constant associated to the energy needed to place a particle in $r_{+}$
($l_{p}=1$).

The total configuration energy is given by the volume integral

\begin{equation}
U_{int} = \int dV \sum_{i=1}^{\bar{N}} \rho \, \mathcal{U}(r_{i}).
\label{totalU}
\end{equation}

In order to give a closed value to Eq. (\ref{totalU}), a mean field (MF) approximation will be made insisting 
that the density of particles is constant throughout the interior of the container, $\rho = \frac{\bar{N}}{V}$. 
Therefore, we approximate Eq.(\ref{totalU}) to 

\begin{equation}
U_{MF} = -\frac{\epsilon \bar{N}^{2}}{V} \int dV \delta(r-r_{+}) 
\label{MFTOTALU}
\end{equation}

and the expression for the mean field energy is

\[ U_{MF} = -\frac{\epsilon \bar{N}^{2} A}{V}, \]

where $A = 4 \pi r^{2}_{+}$. \\

The expression for the interaction energy, as the standard VdW theory states, is
$U_{\mathrm{VdW}} = -\frac{a\bar{N}^{2}}{V}$. In order to have consistency between both results, the proportionality constant $\epsilon$ must be given by 

\begin{equation}
\epsilon = \frac{a}{A} = \frac{k}{8 \pi r^{2}_{+}}.
\label{supE}
\end{equation}

We remind the reader that here the {\it independence condition} has been explicitly used as all calculations have been done on 
the basis the variables $(\bar{N}, V_{t}, T, ...)$ are independent and only at the end they have been mapped to their gravitational 
counterpart, as it can be seen in the right side of Eq. (\ref{supE}).
Following this ``mapping", Eq. (\ref{supE}) can be taken as a sign of the possibility that, in the gravitational theory, 
we might have an energy that relates the curvature of the event horizon with these space--time atoms previously mentioned. 
In particular, the previous equation can be written in terms of the 
gaussian curvature of the horizon, $K$, as $\epsilon = \frac{k}{8\pi^{2}r^{2}_{+}} = \frac{K}{8\pi^{2}}$.
In extension, using the Gauss-Bonnet theorem, one can relate the $a$ parameter with the genus of the horizon. However, the possible
implications of this relation are beyond the scope of this work.

The next step is to use the previous result of MF theory to effectively describe the thermodynamics of the SAdS BH from a 
statistical approach.

\section{Statistical Mechanics}
\label{secIV}

From the previous considerations, we are ready to start proposing a partition function for this gas, provided the 
{\it independence condition}. As $(\bar N,V_{t},T)$ are the independent quantities
, it is appropriate to work in the 
canonical ensemble, whose potential is given by the Helmholtz 
free energy $dF=-SdT-PdV_{t}$. In the context of the extended phase space, the most natural potential to work with 
is the Gibbs potential, as it is easily connected to the mass via a Legendre transformation \cite{Kastor2009}.
 
After these comments, the proposed partition function reads

\begin{equation}
Z_{c} = \frac{\Omega}{\bar{N}!} Z_{0} \times Z_{MF} ,
\label{VdWpartfun}
\end{equation}

where
 
\begin{eqnarray}
Z_{0} &=&\left(\frac{V}{\Lambda^{3}}\right)^{\bar{N}} 
\label{Zgasid} \\
Z_{MF} &=& \exp\left(\frac{\beta a\bar{N}^{2}}{V} \right)
\label{ZMF} \\
\Omega&=&\exp(-\bar{N}).
\label{ZPlus}
\end{eqnarray}

The first term, Eq. (\ref{Zgasid}), represents the ideal gas part within our system.  $\Lambda$ is the 
\textit{thermal De Broglie wavelength} and $\beta$ is the inverse Hawking temperature. In the units here used, $\Lambda$ is 
expressed as $\Lambda = \sqrt{2\pi\beta/m^{*}}$. We point out that the final results should be independent of the mass parameter
$m^{*}$ and, therefore, we are not going to assign any meaning to it. In this sense, $m^{*}$ has to be taken as an instrument 
of the calculation.

The second term, Eq. (\ref{ZMF}), includes the curvature energy which was derived in the previous section. 

The discussion regarding the last term, Eq. (\ref{ZPlus}), will be postponed to section \ref{secVI}. 

With all the previous ingredients, now it is straightforward to construct the Helmholtz free energy potential by taking the 
logarithm of Eq. (\ref{VdWpartfun}),
\begin{equation}
F_{\mathrm{VdW}} = F_{0}-\frac{a\bar{N}^{2}}{V}+\frac{\bar{N}}{\beta},
\label{FreeH}
\end{equation}
where the ideal gas term, $F_{0}$, is given by $F_{0} = \frac{\bar{N}}{\beta}\left(\log(\rho \Lambda^{3})-1\right)$. We
have taken $\bar{N} \gg 1$ in order to use Stirling's approximation. From Eq. (\ref{FreeH}) it is easy to 
recover the eos by 

\begin{align*}
P&=-\frac{\partial F_{\mathrm{VdW}}}{\partial V} = 
-\frac{\partial F_{0}}{\partial V}+\frac{\partial }{\partial V} \left(\frac{a\bar{N}^{2}}{V} \right) \\
&= \frac{\bar{N}}{V\beta} - \frac{a \bar{N}^{2}}{V^{2}},
\end{align*}

as expected.

The next quantity of interest is the entropy, given by the sum of the Sakur--Tetrode equation arising from a VdW gas with
no covolume and from the $\Omega$--term, which can be computed as

\begin{align*}
S_{\mathrm{VdW}}&=-\frac{\partial F_{\mathrm{VdW}}}{\partial T} = -\frac{\partial F_{0}}{\partial T}-\frac{\partial}{\partial T}
\bar{N}T \\
&= \left( \frac{5\bar{N}}{2}-\bar{N} \ln(\rho \Lambda^{3}) \right)-\bar{N} \\
&= \frac{3\bar{N}}{2}-\bar{N} \ln(\rho \Lambda^{3}).
\end{align*} 

In order to match  $S_{\mathrm{VdW}}$ with the Bekenstein--Hawking entropy, $S$, the second term, which depends on the 
combination $\rho \Lambda^{3}$, should be absent. Invoking the {\it independence condition}, the natural choice would be to impose that the density and the thermal wavelength of the $\bar N$ particles at constant $\bar N$ and 
$V_{t}$, in the gravitational theory, satisfy 

\begin{equation}
\rho \Lambda^{3} = 1.
\label{relation}
\end{equation}

The interpretation we give to this constraint will be postponed to the next section, where the extension to
SAdS BHs of any dimension will be developed. 

After imposing Eq. (\ref{relation}) and remembering that $\bar{N} = \frac{N}{6}$ 
we are left with the desired result of

\begin{equation}
S_{\mathrm{VdW}}=\frac{N}{4} = \frac{A}{4}.
\label{BHent}
\end{equation} 

Another quantity of interest is the internal energy of this system, which is given by

\[U_{\mathrm{VdW}} = \frac{3\bar{N}T}{2}-\frac{a\bar{N}^{2}}{V_{t}}. \]

This is the natural result one can expect from the statistical treatment. However, it can be recast into a more suitable expression
for the gravitational case, using Eq. (\ref{vdW}), as

\begin{equation}
U_{\mathrm{VdW}} = \frac{\bar{N}T}{2}+P V_{t}.
\label{uvdw}
\end{equation}  

From this equation we read an internal energy composed of two terms with different origins. The first term is the 
equipartition of 
these $\bar{N}$-particles. It is interesting to point out that eventhough we worked in three dimensions, the equipartition remains
one--dimensional. The second term is associated with a work term provided by the cosmological constant 
interpreted as a vacuum energy. 

\section{Extension to D--dimensions}
\label{secV}
In this section we discuss how the approach here presented can be generalized to $D$ dimensions. 
%
%
In this case, the generalized Smarr relation is given by \cite{Kastor2009}
\begin{equation}
\label{SmarrD}
(D-3)M=(D-2)\frac{\kappa}{8\pi}A-2 \Theta \frac{\lambda}{8\pi},
\end{equation}
where $\Theta = -V_{t}^{(D-1)}$, $V_{t}^{(D-1)}$ stands for the $D-1$ Euclidean volume, $\kappa$ is the surface gravity, 
$T= \frac{\kappa}{2\pi}$ and $P=\frac{\lambda}{8\pi}=\frac{(D-1)(D-2)}{16 \pi l^2}$. 
Even more, following \cite{Altamirano2014},
\begin{equation}
\label{nbarD}
\bar N_{D} = \frac{A^{(D-2)}}{4}\frac{D-2}{D-1},
\end{equation}
where $A^{(D-2)}$ corresponds to the Euclidean area in $D-2$ dimensions.

With these ingredients, Eq. (\ref{SmarrD}) can be stated as
\begin{equation}
\label{vdWD}
P = \frac{\bar N_{D} T}{V_{t}^{(D-1)}}-a_{D}\left(\frac{\bar N_{D}}{V_{t}^{(D-1)}}\right)^2,
\end{equation}
where $a_{D}=\frac{(D-3)}{\pi (D-2)}$.

At this point, the proposed partition function reads
\begin{equation}
Z^{(D)}_{c} = \frac{\Omega^{(D)}_{\bar{N}_{D}}}{\bar{N}_{D}!} Z^{(D)}_{0} \times Z^{(D)}_{MF},
\label{VdWpartfunD}
\end{equation}
where
\begin{eqnarray}
Z^{(D)}_{0} &=&\left(\frac{V^{(D-1)}_{t}}{\Lambda^{D-1}}\right)^{\bar{N}_{D}} 
\label{ZgasidD} \\
Z^{(D)}_{MF} &=& \exp\left(\frac{\beta a_{D}\bar{N}_{D}^{2}}{V^{(D-1)}_{t}} \right)
\label{ZMFD} \\
\Omega^{(D)}_{\bar{N}_{D}}&=&\exp\left(-\frac{D(D-3)}{2(D-2)}\bar{N}_{D}\right).
\label{ZPlusD}
\end{eqnarray}
Note that Eqs. (\ref{ZgasidD}) and (\ref{ZMFD}) are trivial generalizations of Eqs. (\ref{Zgasid}) and (\ref{ZMF}). 

With these tools at hand, the eos given by Eq. (\ref{vdWD}) is easily recovered.
Even more, the statistical entropy can be shown to be
\begin{equation}
S_{D}=(D-1)\frac{\bar {N}_{D}}{2}+\bar {N}_{D}-\frac{D(D-3)}{(D-2)}\frac{\bar{N}_{D}}{2} 
+\bar{N}_{D}\log \left(\rho_{D}\Lambda^{D-1}\right),
\label{EntropyD}
\end{equation}
where $\rho_{D}=\frac{\bar{N}_{D}}{V^{(D-1)}_{t}}$. Moreover, after imposing the corresponding generalization of 
Eq. (\ref{relation}), given by  
\begin{equation}
\label{relationD}
\rho_{D}\Lambda^{D-1}=1
\end{equation}
and using Eq. (\ref{nbarD}), we arrive to
\begin{equation}
S_{D}=\frac{A_{D}}{4},
\end{equation}
which is the desired result.
 
Additionally, the internal energy is given by $$U_{D} = \frac{(D-3)}{2}\bar {N}_{D} T+PV_{t}^{(D-1)}.$$ 
It is interesting to point out that, for a BTZ BH, the internal energy is given only by the work term, 
$PA_{t} = P(\pi r^{2}_{+})$. 

\section{Discussion}
\label{secVI}


\subsection{Gibbs factor and indistinguishability}

A crucial point of the approach here presented is the incorporation of the Gibbs factor, $(\bar{N}_{D}!)^{-1}$. 
One might be tempted to understand this term as a possible sign of indistinguishable particles, following the 
usual statistical theory. Nevertheless, our model does not enlight this fact as we do not know if the complete microscopical theory can support this assumption. Given the statistical treatment here presented, this term must be incorporated to solve for the Gibbs paradox. More precisely, this factor eliminates a divergent term of the form $\bar{N}_{D}\log(\bar{N}_{D})$ (recall $\bar{N} \gg 1$) whilst leaving the term $\log(\rho_{D}\Lambda^{D-1})$. With this being said, the Gibbs term is more an instrument of the calculation rather than the usually understood term. In this sense, we do not adscribe any meaning to $(\bar{N}_{D}!)^{-1}$ but rather follow in the spirit of Ref. \cite{Ashtekar2006}.

\subsection{Heuristic \textit{Stacking} condition}

In this subsection we would like to address the constraint given by $\rho_{D} \Lambda^{D-1}=1$, as it is an unusual requirement 
within the framework of classical statistical mechanics. To start this discussion, let us recall that 
$(\bar{N}_{D},V_{t},T)$ have been considered as independent variables and only at the end of the calculations they depend on
$r_{+}$. 
In fact, we go from the usual one-dimensional specific volume ($v^{(D-1)} = V_{t}^{(D-1)}/\bar{N}_{D}$), which is regularly used 
in the extended phase space approach, to the $(D-1)$-dimensional thermodynamical volume, $V_{t}^{(D-1)}$, which is filled with 
these $\bar{N}_{D}$-particles. It is in this new thermodynamical system that it makes sense to propose a partition function which 
must reproduce a VdW gas. In order to fullfill this goal, a $(D-1)$-{\it thermal de Broglie volume}, given by 
$\Lambda^{D-1}$, is introduced after the integration in the momentum space of the free particle Hamiltonian.

With this clear, in the process of obtaining the entropy, the connection between the gravitational theory and the statistical 
treatment is done when taking the derivative with respect to the temperature leaving $\bar{N}_{D}, V_{t}$ fixed, {\it i. e.}, 
having constant density, $\rho_{D}$. Therefore, the density can now be linked to the expression 
$\rho_{D}= \frac{1}{2l_{p}r_{+}}$, which can be understood as the requirement for fixing $r_{+}$ in the gravitational side. 
Thus, the constraint given by Eq. (\ref{relationD}) can be read as follows: for a fixed horizon radius, $r_{+}$, to construct 
the thermodynamical volume of the BH, the specific volume occupied by these spacetime atoms must equals the 
$(D-1)$-{\it thermal de Broglie volume}, $\Lambda^{D-1}$. This  condition implies that the emergence of the thermodynamical volume, 
$V_{t}$, in the gravitational theory, would be a consequence of the specific volume occupied by the spacetime atoms in the 
statistical theory. This can be visualized as {\it stacking} $\bar{N}_{D}$-boxes of
volume $\Lambda^{D-1}$, where in each box lives one spacetime atom, to construct the thermodynamical volume in a similar way
of building a wall provided the bricks.

Note that the procedure here presented applies for any eos (restricted to the
spherically symmetric case) for arbitrary $D$, where the coefficients that go with higher orders of $\rho_{D}$ are 
independent of the temperature. For a particular example, in $D=4$, the eos for a 
Reissner--N\"{o}rdstrom BH \cite{Kubiznak2012} can be stated as 

\begin{equation}
P_{\mathrm{RN}} = \frac{\bar{N}T}{V_{t}}-\frac{1}{2 \pi}\frac{\bar{N}^2}{V^2_{t}}+\frac{2Q}{\pi}\frac{\bar{N}^4}{V^4_{t}},
\label{RN}
\end{equation}

where we recognize the second coefficient as $a_{2} = \frac{2Q}{\pi}$. If we ask for the entropy, 
the terms proportional to $\rho^{2}_{D}$ and $\rho^{4}_{D}$ are dropped in the derivative as they do not depend on the temperature. 
Therefore, the entropy is still given by 
the Sakur--Tetrode equation minus $\bar{N}$ which, after applying Eq. (\ref{relation}), gives the correct Bekenstein--Hawking 
entropy, as expected. 

Even more, if a VdW gas with covolume $b_{D}$ is considered \cite{Mann2014}, the volume term must be replaced for 
$V^{(D-1)}_{t^{'}} = V^{(D-1)}_{t}-\bar{N}_{D}b_{D}$ and Eq. (\ref{relationD}) must be changed to 
$\bar{N}_{D}\Lambda^{D-1}=V^{(D-1)}_{t^{'}}$. In this case, the interpretation given to Eq. (\ref{relationD}) is
not modified as we are still constructing the available thermodynamical volume by {\it stacking} the spacetime atoms of 
this particular geometry. 

\subsection{Modified counting and propagating gravitational degrees of freedom}

Now we would like to discuss the generalization of Eq. (\ref{ZPlus}) to Eq. (\ref{ZPlusD}). 
The term introduced by means of Eq. (\ref{ZPlus}) in the partition function, given by Eq. (\ref{VdWpartfun}),  
is the one needed to remove the overcounting of states in our entropy, substracting to the original entropy the term 
$D(D-3)\bar{N}_{D}/2(D-2)$. In this sense, the quantity $\left(\Omega^{(D)}_{\bar{N}_{D}}\right)^{-1}\bar{N}_{D}!$ could 
be understood as a modified counting of states which, given the term $D(D-3)$, could be related to the propagating 
gravitational degrees of freedom per spacetime point, given by $D_{\mathit{f}} = D(D-3)$. 


Interestingly, within the approached here employed, the BTZ BH behaves as an ideal gas ($a_{3}=0$) and its entropy 
can be computed without any reference to Eq. (\ref{ZPlus}) because in $D=3$ it does not contribute 
($\Omega^{(3)}_{\bar{N}_{3}}=1$). This is related to the fact that, in three dimensions, there are no propagating 
gravitational degrees of freedom. 


\section{Concluding remarks}
\label{secVII}

We have constructed a heuristic picture of a SAdS black hole as a Van der Waals gas of spacetime atoms, considering 
a spherical container of fixed volume and constant Hawking temperature where these spacetime atoms live. A statistical 
approach using the Boltzmann distribution has been performed in order to reproduce the gravitational thermodynamics, with
emphasis on both the Smarr (Van der Waals) law and the Bekenstein--Hawking entropy. The procedure here presented
applies for $D\ge 3$ SAdS black holes and also for the charged $D=4$ case.
In this sense, we coincide with Ashtekar's comment \cite{Ashtekar2006} by replacing {\it punctures} by
{\it spacetime atoms}: ``at the end of the calculation,
one may try to reinterpret the result by constructing a heuristic picture of the black hole horizon as a gas of punctures. 
One can then conclude that one would reproduce the result of the calculation by using
Boltzmann statistics for the hypothetical puncture particles".

To end the discussion, we would like to synthesize the main lessons that can be extracted from this {\it dual}
picture of a SAdS black hole. By considering a microscopic entity, namely the spacetime atoms, the thermodynamics of the SAdS
black hole can be obtained, after some heuristic considerations. The statistical picture sheds some light on the
behaviour and some unusual features of these dof, via the heuristic constraints that must be applied to succesfully reproduce the
Bekenstein--Hawking entropy. In particular, from the {\it stacking condition,} two main lessons can be learned. First,
the apparition of an intrinsic ``statistical" feature of the spacetime atoms, enclosed in their De-Broglie wavelength, manifests
itself in the corresponding gravitational theory by the emergence of the thermodynamical volume of the EPS approach,
which is a new perspective for this quantity. Second, the De-Broglie wavelength might serve as an indication of the
existence of a corpuscular limit of a full microscopical theory which should explain the constraint from first principles.
In another instance, the {\it modified counting and propagating gravitational dof} requirement reveals a new perspective on the
spacetime atoms as Eq. (\ref{ZPlusD}) endow them with some dof that affect the possible states they can access, following the
statistical perspective. An interesting problem would be to find how does this term arises naturally from a microscopic theory for
the spacetime atoms.

As a final comment we point out that some relations between our approach and both Euclidean path integral and extended phase
space methods must exist. In particular, possible links between the Euclidean partition function and the one
here proposed could enlight, at least heuristically, some microscopical components of gravity. 

\section{Acknowledgments} The authors would like to acknowledge J. F. M\'endez, F. D. Villalba, N. Morales--Dur\'an, G. T\'ellez  and 
R. Mart\'{\i}n--Landrove for fruitful discussions. The authors acknowledge the support from 
the Faculty of Science and  Vicerrector\'{\i}a de Investigaciones of Universidad de Los Andes, Bogot\'a, Colombia.

\end{document}